\documentstyle[preprint,aps]{revtex}
\begin{document}
\draft

\title{E2/M1 ratio from the Mainz $p(\vec \gamma, p)\pi^0$ data}
\maketitle

In a recent Letter~[1], Beck {\it et al.} have determined the E2/M1 ratio
using differential cross section and photon asymmetry data from the 
$p(\vec \gamma, p)\pi^0$ reaction at the Mainz Microtron MAMI. After 
the correction of a sign error, Eqs.~(4) and (6) of [1] read
\begin{eqnarray}
A_{||} \; & = & \; | E_{0+} |^2 \; + \; | 3E_{1+} - M_{1+} + M_{1-} |^2,  \\
C_{||} \; & = & \; 12 {\rm Re} \left[ E_{1+} \left( M_{1+} - M_{1-} 
\right)^* \right],
\end{eqnarray}
for the constant and $\cos^2 (\theta)$ terms in the parallel ($||$) component
of the differential cross section. 
 
In eq.~(7) of [1], the following association is made 
\begin{equation}
R \; = \; {    { {\rm Re} \left( E_{1+} M_{1+}^* \right) } \over
               { | M_{1+} |^2 }    }  \; \simeq \; {1\over 12} 
               { {C_{||} }\over {A_{||} } } ,
\end{equation}
between the ratio of $C_{||}$ and $A_{||}$ coefficients, 
and the ratio of multipoles giving the E2/M1 ratio at resonance.
At the resonant point, a simplified expression is given: 
\begin{equation}
R \; = \; {    { {\rm Im} E^{3/2}_{1+} } \over
               { {\rm Im} M^{3/2}_{1+} }    } \; = \; R_{\rm EM}.
\end{equation}

The authors of [1] note that the ratio $C_{||} / ( 12 A_{||} )$ has a 
constant value of $-2.5\%$ across the resonance. As a result, they quote
$( -2.5\pm 0.2\pm 0.2)\%$ for the E2/M1 ratio, the systematic error coming
from the limited angular efficiency of their detector and ignored isospin
1/2 contributions. 

If we neglect, in our Eqs.~(1) and (2), all contributions apart from those
involving $|M_{1+}|^2$ and ${\rm Re}(E_{1+} M_{1+}^*)$, we actually have
\begin{equation}
{1\over 12} { {C_{||} }\over {A_{||} } } \; \simeq \;
{ {R_{\rm EM} }\over {1 \; - \; 6 R_{\rm EM} } }  ,
\end{equation}
at the resonant point. 
Neglect of the $R_{\rm EM}$ term in the denominator results in an error
of about 17\%  for $R_{\rm EM}$, which is more than double the systematic 
error quoted in [1]. 

Using our multipole amplitudes~[2], we find this effect is reduced due to
a cancellation between Im $M_{1-}$ and Im $E_{1+}$ in Eq.~(1). The extent
to which this applies to the result of [1] is unclear, as we find~[2] a 
different value, $(-1.5\pm 0.5)\%$, for the E2/M1 ratio.  

We suggest that the systematic error on the E2/M1 ratio, extracted in [1],
should be enlarged. We also note that, using the corrected signs
in our Eq.~(1), the E2/M1 ratio is increased in magnitude.  
This actually {\it worsens} the agreement between our value and the value 
found in [1].

\vskip .5cm

R.L. Workman

Department of Physics, 

Virginia Polytechnic Institute and State University, 

Blacksburg, VA 24061

\vskip .5cm

PACS numbers: 11.80.Et, 13.60.Rj, 25.20.Lj
 
\vskip .5cm

[1] R. Beck et al., 
Phys. Rev. Lett. {\bf 78}, 606 (1997). 

[2] Our multipole amplitudes, determined in a fit to low-energy data,
are accessible through the SAID program (the solution is W500). Telnet
to clsaid.phys.vt.edu with the userid: said.
\vfill

\end{document}